# Skyrmion-mediated Nonvolatile Ternary Memory

[1]Md Mahadi Rajib, [3]Namita Bindal, [3]Ravish Kumar Raj, [3]Brajesh Kumar Kaushik and [1,2]Jayasimha Atulasimha

[1]Department of Mechanical and Nuclear Engineering, Virginia Commonwealth University, Richmond, VA 23284, USA

[2]Department of Electrical and Computer Engineering, Virginia Commonwealth University, Richmond, VA 23284, USA

[3]Department of Electronics and Communication Engineering, Indian Institute of Technology Roorkee, Roorkee, Uttarakhand 247667, India

**Abstract**

Multistate memory systems have the ability to store and process more data in the same physical space as binary memory systems, making them a potential alternative to existing binary memory systems. In the past, it has been demonstrated that voltage-controlled magnetic anisotropy (VCMA) based writing is highly energy-efficient compared to other writing methods used in non-volatile nano-magnetic binary memory systems. In this study, we introduce a new, VCMA-based and skyrmion-mediated non-volatile ternary memory system using a perpendicular magnetic tunnel junction (p-MTJ) in the presence of room temperature thermal perturbation. We have also shown that ternary states {-1, 0, +1} can be implemented with three magnetoresistance values obtained from a p-MTJ corresponding to ferromagnetic up, down, and skyrmion state, with 99% switching probability in the presence of room temperature thermal noise in an energy-efficient way, requiring ~3 fJ energy on an average for each switching operation. Additionally, we show that our proposed ternary memory demonstrates an improvement in area and energy by at least 2X and ~60X respectively, compared to state-of-the-art spin-transfer torque (STT)-based non-volatile magnetic multistate memories. Furthermore, these three states can be potentially utilized for energy-efficient, high-density in-memory quantized deep neural network implementation.

**Introduction**

In the field of computer technology, CMOS-based two-state memory is widely used. When evaluating memory systems, key considerations include writing and reading speed, reliability, endurance, non-volatility, high density, and energy-efficiency [1]. As CMOS-based two-state memory systems are volatile and are reaching their limits for high-density implementation, researchers are searching for alternatives. Currently, flash memory is the most advanced non-volatile option available, however, it has an endurance problem [1-3]. Other potential options that are still being researched include resistive random-access-memory (RRAM) [4] phase change memory (PCM) [5] magnetoresistive random-access memory (MRAM) [6, 7] and ferroelectric random-access memory (FeRAM) [8, 9] with MRAM being the most promising [1, 10]. MRAM devices are made up of nanomagnets, where the "up" and "down" states typically represent the "0" and "1" bits in a p-MTJ as shown in Fig. 1(a). There are two main methods to write these bits, a current-dependent method [11-14] and an electric field-dependent method [15-19]. The current-dependent approach typically results in high energy dissipation when switched with STT [11,12,20]. Using spin-orbit torque (SOT) could potentially improve energy-efficiency, however, it requires a three terminal device geometry [13]. On the other hand, electric field-based write approach can be more energy-efficient whether mediated through strain [21-27] or voltage control of magnetic anisotropy [17, 28, 29] or other methods such as electrical switching of polarization coupled to antiferromagnetic state [30]. In particular, MTJs switched with strain can potentially be scaled to switch at < 1 fJ/bit [31] while voltage-controlled magnetic anisotropy-based switching, a type of electric field-dependent method, requires only a few fJ [32, 33] of energy per switching operation compared to 100 fJ using STT switching (a current-dependent method) [34]. Along with energy-efficiency high density of RAMs is also desirable, so a multi-state approach is more practical. In the past, multistate MRAM has been proposed using STT [35] and SOT [36], but these methods result in decreased density because they involve connecting MTJs in parallel or series. Three-terminal devices like spin-orbit torque magnetic random-access-memory (SOT-MRAM) require more space, and multibit spin-transfer torque magnetic random-access-memory (STT-MRAM) requires even more energy than its two-state counterpart. Therefore, there is a need to implement a multibit memory system in a single MTJ with two terminals in an energy-efficient way. Furthermore, ternary memory can be of utility for highly quantified deep neural networks.

In this study, a VCMA-based and skyrmion-mediated ternary memory system using a single MTJ with two terminals is presented. In addition to the standard "up" and "down" ferromagnetic states of the two-state MRAM system, the third state in the proposed ternary memory system is a skyrmion state. In our system, the skyrmion state exhibits a conductance roughly equal to the average of the highest ("up" ferromagnetic state) and lowest ("down" ferromagnetic state) conductance values which is discussed in the results and discussions section in details. As a result, the skyrmion-mediated memory can be viewed as an almost balanced ternary memory, with resistance states of +1 (ferromagnetic up), ~ 0 (skyrmion), and -1 (ferromagnetic down). Here we note that the magnetization of the reference layer is considered to be pinned in the upward direction. Skyrmions are topologically protected states that offer nanoscale size, high velocity, and low depinning current density [37-39]. These spin textures are usually used in racetrack memory devices [38, 40, 41], but in this study, we propose the usage of a skyrmion in a confined structure like the MTJ. Our group has theoretically shown that switching between the ferromagnetic and skyrmion states in the MTJ's free layer can be done using VCMA at 0 K [42], and also experimentally shown that skyrmions can be created and annihilated in a thin film using VCMA [43]. However, the creation and annihilation of skyrmions at room temperature in a confined structure has not been reported yet. This study theoretically demonstrates that a dense and energy-efficient ternary memory can be implemented by alternating between the three states (ferromagnetic up, ferromagnetic down, and skyrmion as shown in Fig. 1(b)) using VCMA in a patterned 100 nm nanodot in the presence of thermal noise at room temperature. We have shown in other work that such skyrmion based VCMA two-state memory can be scaled to < 50 nm [44,45] which suggests that this approach also has the potential for scalable ternary memory.

The proposed three state memory system can also be used for implementing neuromorphic computing devices, particularly as quantized synaptic weights for deep neural networks (DNN). The demand for neuromorphic computing is growing rapidly due to its ability to handle training and inference tasks in energy-limited environments like edge devices [46-48]. In DNN, the aspect that requires the most time and energy is the vector-matrix multiplication operation [49]. In-memory computing offers a more efficient and low-energy solution to this challenge [48]. The previously mentioned non-volatile memory types, such as magnetoresistive memory [50, 51, 52], flash memory [53], resistive memory [54,55], and phase-change memory [56, 57], have been also shown to be useful in performing multiply-accumulate operations commonly found in artificial neural networks [58]. Of these, magnetoresistive memory has been extensively researched for its high endurance, low energy requirements for reading and writing, non-volatility, and compatibility with CMOS technology [51, 59]. Recently, it has been demonstrated that use of three or five-state domain wall based quantized synaptic weights in deep neural networks can perform recognition tasks with an accuracy comparable to that of floating-point precision synaptic weight based DNNs with low energy consumption [51]. However, writing multiple states in domain wall-based racetrack devices requires a lot of space. In this study, we also discuss that our proposed three-state memory system, implemented in a two-terminal MTJ, allows for the potential implementation of quantized synaptic weights with a comparable energy consumption to domain wall-based DNNs, but with significantly less space requirements.

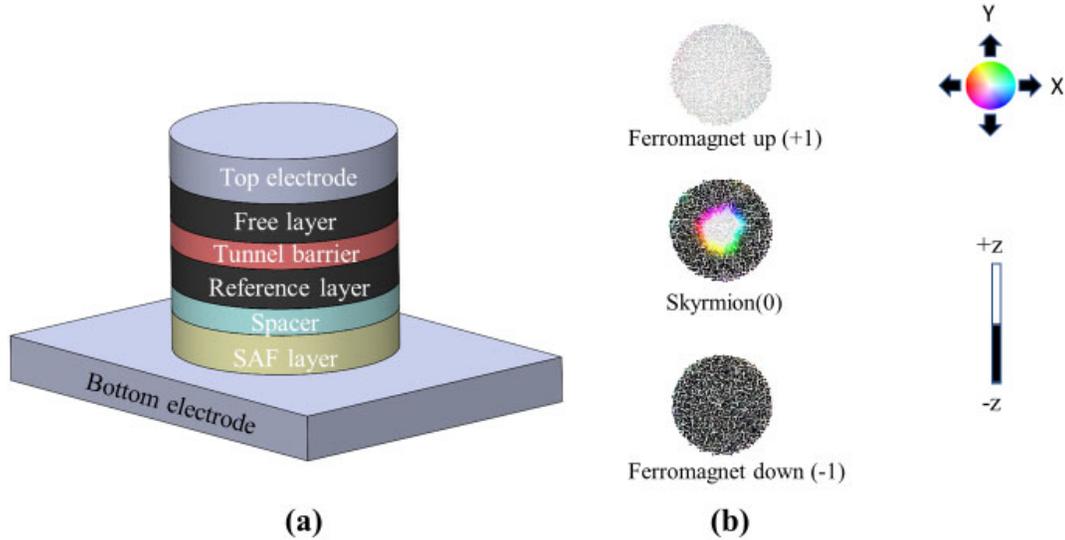

Figure 1. (a) Schematic of a magnetic tunnel junction with different constituent layers. (b) Ferromagnetic up, skyrmion and ferromagnetic down states represent +1, ~0, -1 states, respectively of the proposed ternary memory.

**Results and Discussion**

There are three different switching operations that are primarily carried out: switching from a ferromagnet (+1/-1) to a skyrmion state (0), switching from a skyrmion (0) to a ferromagnet state (+1/-1), and switching from one ferromagnet (+1/-1) to another (-1/+1). Here we note that the reference layer is magnetized in the upward direction in the p-MTJ. Therefore, the ferromagnetic up and down state has the highest and lowest conductance, respectively, while the skyrmion state has an intermediate conductance between the highest and lowest [60], which is nearly half of the combined conductance in the case of skyrmions with equal number of up and down spins. This allows the execution of all six interconversions needed to implement a three-state memory. However, as shown in Fig. 2, eight situations occur while performing switching operations.

In case A, switching from ferromagnetic up (+1) to skyrmion state (0) is performed. For performing this switching operation, the ferromagnetic up state is relaxed for 1 ns and perpendicular magnetic anisotropy (PMA) energy is reduced from 2250 µJ/m$^2$ to 1575 µJ/m$^2$ in 0.1 ns by applying a +1 V pulse. Here, we note that it has been previously experimentally observed that positive voltage pulse reduces PMA while negative voltage pulse increases PMA [43, 61]. When the PMA is reduced, the presence of Dzyaloshinskii-Moriya interaction (DMI) field in the ferromagnetic free layer creates a skyrmion state of polarity -1. Here we note that the skyrmion polarity can be defined as p = [m$_z$ (r = ∞) − m$_z$ (r = 0)]/2 [62]. Therefore, skyrmion with boundary (core) pointing down (up) and core (boundary) pointing up(down) has polarity -1 (+1). We observe that when the PMA is reduced from the ferromagnetic up state, a skyrmion of polarity -1 is created because the boundary spins tilt in the opposite direction from the core spins and the core spins are in their starting spin orientation. The voltage pulse is applied for 0.3 ns and subsequently withdrawn in 0.1 ns to restore the initial PMA of 2250 µJ/m$^2$. We observe that the skyrmion state is stabilized after the withdrawal of the voltage pulse. Thus, by applying and subsequently withdrawing a +1 V pulse, a skyrmion state (0) can be created and stabilized starting from a ferromagnetic up state (+1).

In case B, we start from a ferromagnetic down state (-1) and when PMA is reduced in 0.1 ns by applying a +1 V pulse, a skyrmion of polarity +1 is created. The skyrmion produced in this instance has the opposite polarity of the one produced in case A because we start from the opposite ferromagnetic state, the ferromagnetic down state. Thus, switching from ferromagnetic down to skyrmion state can be accomplished in a single step similar to case A.

In case C, ferromagnetic up state is switched to ferromagnetic down state. This is a two-step operation. First, a +1 V pulse is applied to reduce the PMA and create a skyrmion state. The created skyrmion has core oriented in the upward direction and boundary in the downward direction. When a -1V pulse is applied to increase the PMA from 2250 $\mu J/m^2$ to 2925 $\mu J/m^2$ in 0.1 ns, the skyrmion annihilates to the ferromagnetic down state following the skyrmion's downward oriented boundary spins. This switching is deterministic since ferromagnetic up state is switched to ferromagnetic down state through an intermediate skyrmion state which has downward oriented boundary spin and the skyrmion annihilates to ferromagnetic down state following the boundary spins' orientation. We can see from Fig. 2 that the ferromagnetic down state is stable after the PMA is restored by withdrawing the +1V pulse.

Similarly, in case D, ferromagnetic down state is switched to ferromagnetic up state through an intermediate skyrmion state. In this case, starting from a ferromagnetic down state, an intermediate skyrmion with boundary spin in the upward direction and core in the downward direction is created by applying +1V pulse and subsequently the skyrmion is annihilated to ferromagnetic up state by applying -1V pulse. Therefore, writing -1/+1 to +1/-1 is a two-step operation where a positive and a negative voltage pulse is applied sequentially.

Writing from skyrmion to ferromagnet can be one step/three step operation. We have seen in case C and case D that while switching from one ferromagnetic to other ferromagnetic state through an intermediate skyrmion state, the skyrmion state annihilates to ferromagnetic state following the skyrmion state's boundary spin orientation. Writing a desired ferromagnetic state from a skyrmion state can be a one-step or three-step operation depending on the skyrmion polarity because we can't tell the skyrmion's polarity from its magnetoresistance value. For example, if the skyrmion has upward oriented boundary spin then it will annihilate to a ferromagnetic up state (case E). But, if the desired state is the downward ferromagnetic state then the ferromagnetic up state is required to be switched by following the similar operations performed in case C and case D.

In case E, we can see that skyrmion with boundary spin in the upward direction can be switched to a ferromagnetic up state by applying a -1V pulse. If the desired state is ferromagnetic down state then a sequential application of +1 V and -1V pulse will switch the ferromagnetic up state to ferromagnetic down state requiring a three- step operation overall which is shown in case F. If the initial state is a skyrmion state with downward boundary spin and a ferromagnetic down state needs to be written, then the application of a +1V pulse completes the switching as shown in case G. If the desired state is ferromagnetic up state then a sequential application of +1V and -1V pulse will be required to switch the ferromagnetic down state to ferromagnetic up state as shown in case H.

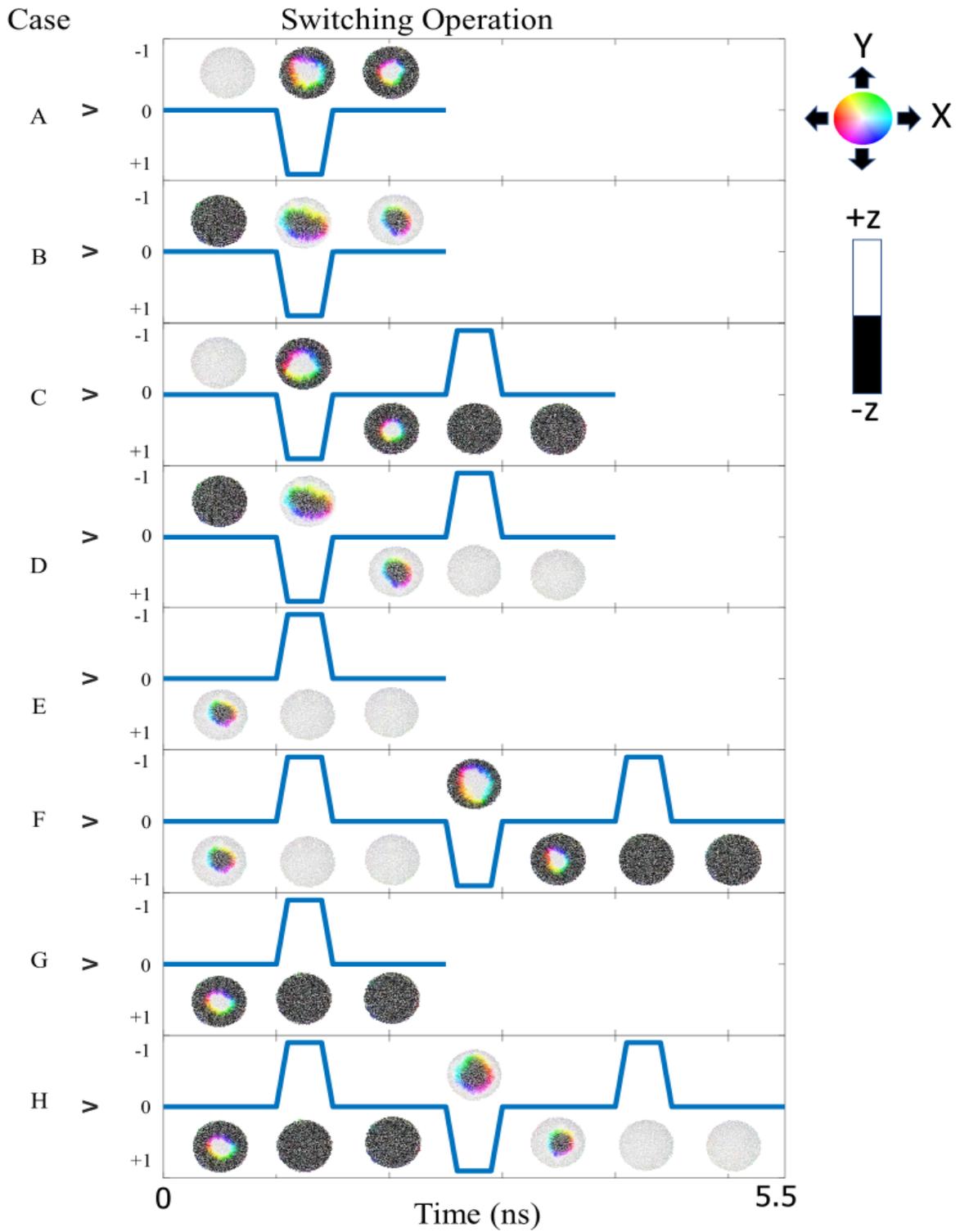

Figure 2. Applied voltage vs time for different switching cases. (The magnetization states for different voltage pulses are shown correspondingly)

We ran simulations for 1000 times to study the switching percentage in each of the eight cases in the presence of room temperature thermal perturbation, and the results are shown in Table 1. On average, switching occurred in ~99% of the cases.

**Energy dissipation:**

We calculated the energy required for each switching operation considering all of the reading and writing operations involved. For calculating the energy dissipation, the generalized conductance of the MTJ is considered to have two conductance $G_1$ and $G_2$ in parallel as shown in Fig. 3. Therefore, the net conductance of the MTJ is calculated as follows:

$$G_{MTJ} = G_P \left(\frac{A_{RL}-A_{FL/RL}}{A_{RL}}\right) + G_{AP} \frac{A_{FL/RL}}{A_{RL}} \qquad (1)$$

where $A_{RL}$ and $A_{FL/RL}$ are the area of reference layer (RL) and the area of the free layer (FL) representing the antiparallel region with respect to the reference layer, respectively.

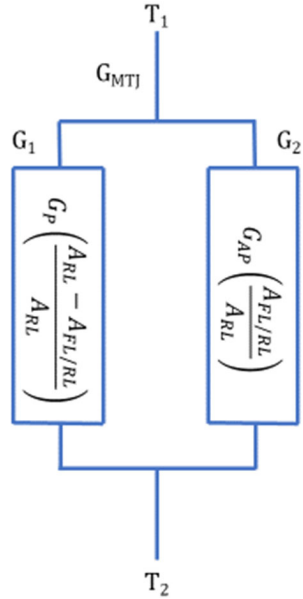

Figure 3. Generalized model of memory state of MTJ

According to the equation 1, the conductance of the parallel and antiparallel states in the MTJ are $G_P$ and $G_{AP}$, respectively. However, when a skyrmion is formed in the FL region with nearly an equal number of spins in the upward and downward direction, it can be assumed that ~50% area of FL is antiparallel to the RL, meaning $A_{FL/RL} \cong 1/2\, A_{RL}$. Therefore, from equation (1) $G_{MTJ} \cong \frac{G_P+G_{AP}}{2}$, which indicates that the conductance of skyrmion state is nearly the average of the conductance of parallel and anti-parallel state. We consider 891.27 Ω resistance [63] for the parallel orientation of the reference and free layers of the p-MTJ. Considering the TMR ratio 249% [63], therefore, the antiparallel and skyrmion state has 3110.53 Ω and ~2000.90 Ω resistance respectively.

While implementing skyrmion-mediated ternary memory, energy is dissipated during both the reading and writing processes that involves $I_{read}^2 R_{MTJ} t_{read}$ and $\frac{1}{2} CV_{write}^2$ loss, respectively. To read the states, a 600 mV pulse is applied for 1 ns, while different combinations of read and write pulses are applied to write different states. For example, in case A, the initial and final states are read as ferromagnetic up state, and the skyrmion state respectively and the energy dissipated for reading is 0.975fJ and 0.886fJ, respectively, thus the total energy dissipated for reading is 1.861fJ. To write the skyrmion state, a +1V is applied for 0.5 ns, resulting in 0.337fJ of switching energy. Therefore, operating case A requires a total of 2.198 fJ of energy. Table 1 shows the energy required for switching from one state to another for all the eight cases. On average, the energy dissipated in writing the three states is 3.19 fJ.

Table 1. Switching percentage and energy dissipation for performing different switching operations

| Case | Switching percentage | Write energy (fJ) | Read energy (fJ) | Total energy (fJ) |
|------|----------------------|-------------------|------------------|-------------------|
| Case A | 98.70 | 0.337 | 1.861 | 2.198 |
| Case B | 98.70 | 0.337 | 1.739 | 2.076 |
| Case C | 98.80 | 1.011 | 2.714 | 3.725 |
| Case D | 99.00 | 1.011 | 2.714 | 3.725 |
| Case E | 100 | 0.337 | 1.861 | 2.198 |
| Case F | 99.10 | 1.685 | 3.600 | 5.285 |
| Case G | 100 | 0.337 | 1.739 | 2.076 |
| Case H | 98.9 | 1.685 | 3.600 | 5.285 |

Previously, nonvolatile multistate memories (3 or 4 states) were proposed using STT or SOT current, and these memories were created by arranging the MTJs in series or parallel configurations [35, 36, 64]. Therefore, implementing multistate memories with STT and SOT current requires at least twice the area of our proposed multistate memory, as shown in Table 2. It is worth noting that in STT/SOT-based multistate memories, two consecutive switching operations are performed to write a single state, resulting in latency issues [64]. Considering the energy required for performing a switching operation with STT current in a single MTJ with a 100 nm diameter, which is the same as the dimension of our proposed device (100 fJ) [65], writing a state in a multistate STT-MRAM would require ~200fJ of energy. On the other hand, although SOT memories require less energy than STT for an MTJ of similar diameter [66] for writing each bit, the requirement of three terminals reduces the density. Thus, our proposed ternary memory achieves at least a 2X improvement in footprint and a ~60X improvement in energy-efficiency compared to STT-based multistate memory.

With an appropriate choice of material parameters, there is still the possibility of reducing the write error rate in the presence of room temperature thermal noise [44,45]. However, the proposed three-state memory system, which achieves ~99% switching accuracy, has the potential to be utilized as quantized synaptic weights for deep neural networks. The previous idea to use three-state domain wall-based quantized synaptic weights achieved similar levels of accuracy during testing as floating-point precision synaptic weights [51]. However, the implementation of the three states requires a large racetrack (600nm× 60nm) [51], and the use of five terminals reduces the density significantly. Our proposed three-state memory system, which uses a single two-terminal MTJ, provides a way to implement quantized synaptic weights that consume similar amounts of energy as domain wall-based DNNs, but with significantly less space requirement (~4.5 times less space for the MTJ alone) as shown in Table 2.

Table 2. Comparison of different multistate systems

|  | STT-based two state memory | STT-based multistate memory | DW-based stochastic multistate | Skyrmion-mediated ternary memory |
|---|---|---|---|---|
| Number of terminals | 2 | 2 | ≥4 [51, 64] | 2 |
| Area | X | 2X [64] | ~4.5X [51] | X |
| Energy | 100 fJ [65] | ~200 fJ | ~3 fJ [51] | ~3 fJ |

In summary, we showed that a novel skyrmion-mediated ternary memory can be implemented in a p-MTJ in the presence of room temperature thermal noise. We also showed that our proposed ternary memory achieves at least a 2X improvement in footprint and a ~60X improvement in energy-efficiency respectively compared to STT-based multistate memory. By utilizing energy-efficient VCMA switching mechanism and employing a two-terminal MTJ device, our proposed memory design allows for the implementation of three distinct states within a single MTJ, thereby increasing both the cell density and the density of the peripheral circuit. Previously, it has also been theoretically shown that skyrmions can be scaled down to ~20 nm in a circular patterned nanodot [45], which could lead to even greater memory density in the future. Furthermore, three state synapses can be built with comparable energy costs and reduced space requirements (~4.5 times less area) compared to domain wall-based quantized DNNs.

**Methods**

We simulate the magnetization dynamics of the free layer of the circular shaped p-MTJs with a diameter of 100 nm which are discretized into 50×50×1 cells to observe the switching between three states and evaluate the switching probability of each switching case. The simulation uses the MuMAX3 program [67] to solve the magnetization dynamics based on Landau-Lifshitz-Gilbert (LLG) equation that is defined as follows:

$$\frac{\partial \vec{m}}{\partial t} = \left(\frac{-\gamma}{1+\alpha^2}\right)\left[\vec{m} \times \vec{B}_{eff} + \alpha\{\vec{m} \times (\vec{m} \times \vec{B}_{eff})\}\right] \quad (2)$$

where $\alpha$ and $\gamma$ denote the Gilbert damping coefficient and gyromagnetic ratio, respectively. $\vec{m}$ indicates the normalized magnetization vector with components $m_x$, $m_y$, and $m_z$ along $x$, $y$, and $z$ direction, respectively, which is obtained by normalizing the magnetization vector ($\vec{M}$) with respect to saturation magnetization ($M_s$). The circular shaped free layers are divided into grids with dimensions of 2nm×2nm×1.5nm, which are much smaller than the exchange length ($\sqrt{\frac{2A_{ex}}{\mu_0 M_S^2}}$). In the LLG equation $\vec{B}_{eff}$ is the effective magnetic field having the following components:

$$\vec{B}_{eff} = \vec{B}_{demag} + \vec{B}_{exchange} + \vec{B}_{DM} + \vec{B}_{anis} + \vec{B}_{thermal} \quad (3)$$

In equation (3), $\vec{B}_{demag}$ represents the effective field due to demagnetization energy and $\vec{B}_{exchange}$ denotes the Heisenberg exchange interaction respectively.

$\vec{B}_{DM}$ is the effective field due to DMI, which is expressed as follows:

$$\vec{B}_{DM} = \frac{2D}{M_s}\left(\frac{\partial m_z}{\partial x}, \frac{\partial m_z}{\partial y}, -\frac{\partial m_x}{\partial x} - \frac{\partial m_y}{\partial x}\right) \quad (4)$$

Where $D$ represents the DMI constant.

The perpendicular anisotropy ($\vec{B}_{anis}$) is expressed by the following equation:

$$\vec{B}_{anis} = \frac{2K_{u1}}{M_s}(\vec{u}.\vec{m})\vec{u} \quad (5)$$

where $K_{u1}$ and $\vec{u}$ represent the first order uniaxial anisotropy constant and unit vector in the anisotropy direction respectively.

The following equation is used to introduce the thermal field:

$$\vec{B}_{thermal} = \vec{\eta}(step)\sqrt{\frac{2\alpha k_B T}{M_s \gamma \Delta V \Delta t}} \quad (6)$$

where T is the temperature (K), $\Delta V$ is the cell volume, $k_B$ is the Boltzmann constant, $\Delta t$ is time step and $\vec{\eta}$ (step) is a random vector from a standard normal distribution. Here, we note that the random vector $\vec{\eta}$ is independent (uncorrelated) for each of the three Cartesian coordinates and is generated at every time step.

The parameters for Co based magnetic materials are used for the simulation of magnetization dynamics of the skyrmions that are given in Table 3.

Table 3. Material properties

| | |
|---|---|
| Saturation magnetization ($M_s$) | $1.3 \times 10^6$ A/m [68] |
| Exchange stiffness ($A_{ex}$) | $20 \times 10^{-12}$ J/m [69] |
| DMI | $3.0 \times 10^{-3}$ J/m$^2$ [70] |
| Thickness | 1.5 nm |
| Damping coefficient | 0.1 [71] |
| Perpendicular magnetic anisotropy | 2250 µJ/m$^2$ [72] |
| VCMA coefficient | 675 fJ/Vm [72] |

**Acknowledgements**

M. M. R. and J. A. would like to acknowledge US National Science Foundation CISE SHF Small grant # 1909030. N.B., R. K. R. and B. K. K would like to acknowledge Science and Engineering Research Board (SERB), Department of Science and Technology, Government of India under Grant CRG/2019/004551 for providing the funding to carry out the research work.

**Competing interests**

The authors declare no competing interests.